# Direct excitation of a single quantum dot with cavity-SPDC photons

UTTAM PAUDEL,[1,6,*] JIA JUN WONG,[2] MICHAEL GOGGIN,[3] PAUL G. KWIAT,[2] ALLAN S. BRACKER,[4] MICHAEL YAKES,[4] DANIEL GAMMON,[4] AND DUNCAN G. STEEL[1,5]

[1]*Department of Physics, University of Michigan, Ann Arbor, Michigan, 48109, USA*
[2]*Department of Physics, University of Illinois, Urbana Champaign, Illinois 61801, USA*
[3]*Department of Physics, Truman State University, Kirksville, Missouri, 63501, USA*
[4]*Naval Research Laboratory, Washington, DC, 20375, USA*
[5]*EECS Department, University of Michigan, Ann Arbor, Michigan, 48109, USA*
[6] *Present address: The Aerospace Corporation, El Segundo, CA, 90245, USA*
\* *upaudel@umich.edu*

**Abstract**

The ability to generate mode-engineered single photons to interface with disparate quantum systems is of importance for building a quantum network. Here we report on the generation of a pulsed, heralded single photon source with a sub-GHz spectral bandwidth that couples to indium arsenide quantum dots centered at 942 nm. The source is built with a type-II PPKTP down-conversion crystal embedded in a semi-confocal optical cavity and pumped with a 76 MHz repetition rate pulsed laser to emit collinear, polarization-correlated photon pairs resonant with a single quantum dot. In order to demonstrate direct coupling, we use the mode-engineered cavity-SPDC single-photon source to resonantly excite an isolated single quantum dot. © 2019 Optical Society of America

## 1. Introduction

Exploiting quantum coherence and entanglement for secure long distance information sharing has garnished much interest over the past few decades [1]. However, most quantum states are hard to generate on-demand, are fragile to transmit over long distance, and are prone to decoherence, making long-distance transfer of quantum information a formidable engineering feat. A quantum repeater is a small module for a large quantum network that can store information at local nodes, perform specific logic operations, purify quantum states with error correction codes, and teleport the states to the next module [2]. Such quantum repeaters form intermediate links between nodes, allowing processing and transmission of quantum information over a long distance without losing its fragile quantum properties.

Information between two quantum systems can be exchanged through several different ways. By bringing the two matter qubits in proximity, they can be entangled with each other through the local Coulomb interaction [3-4]. Alternatively, by performing Hong-Ou-Mandel (HOM) type interference between the flying qubits emitted by the two matter nodes of interest, one can interface distant nodes [5-7]. If the emitted photons are entangled with the nodes, the distant nodes can be entangled with each other with a Bell state's analysis [5].

Similarly, there exists a third class of protocols that allows information transfer between two distant systems through the direct absorption of single photons [8-17]. Cirac et al. developed a protocol where two atoms of interest for quantum linkage are embedded inside two high-Q cavities [13]. One of the atoms is optically manipulated using lasers and is prepared in a desired quantum superposition state. In the process, the atom emits a packet of photons with the internal state of the atom mapped on to the photon. The photon packet propagates to a nearby atomic site through a waveguide or a transmission channel and gets absorbed by the second atom. Through the absorption process, theoretical studies show the internal state of the first atom can be mapped to the second atom with unit probability, creating a direct link for the exchange of quantum information between two spatially separated atoms [13]. Several experimental efforts have reported progress towards this goal [11, 12, 14-17].

A realistic quantum network could be built out of several different systems, each with different electronic and optical properties. To envision such a hybrid network that could utilize the best features of each system, one needs to build a source of highly flexible flying qubits that can form links between two disparate systems [18, 19]. Among various flying qubit sources, spontaneous parametric down-conversion (SPDC) is a strong contender to form a flexible photonic link in a quantum network, as it can generate a pair of entangled photons at room temperature with engineerable optical properties [20, 21]. With appropriate phase matching conditions, a type-II SPDC source can be designed to emit either degenerate or highly non-degenerate photon pairs that are wavelength-tunable by several hundred nanometers [21]. This allows one to link two systems with very different optical properties. In addition, by placing a down-conversion crystal inside an optical cavity, it is shown that the emitted photons' temporal and spectral properties can be modified as desired while enhancing the count rates of the down-conversion photons [16, 19, 22, 23]. Such large wavelength tunability, along with the customizable spectrum, makes cavity-SPDC an excellent system with which to construct highly flexible entangled photon sources.

In this Article, we report on the realization of a cavity-SPDC source that matches its spectral and temporal properties with that of single photons emitted by InAs/GaAs quantum dots (QDs). For maximum HOM interference between the SPDC source and QD, the spectral and temporal profile of the SPDC photon has to match closely with that of the QD photons [5-7]. Thus, the cavity-SPDC source was designed to generate a decaying exponential to match the radiative decay of the QD for application to a measurement-based entanglement using the HOM interference effect [5-7]. In addition, we demonstrate coupling of single-photons emitted by the cavity-SPDC source with an isolated single QD through direct excitation process. We note that this is in contrast to a rising exponential that is needed to optimize absorption [17, 24-26].

## 2. Mode engineered cavity-SPDC source

We use an X-cut, 5 mm long periodically-poled KTP (PPKTP) crystal for type-II down-conversion and a 2 mm long KTP crystal with a curved surface to compensate the birefringence mismatch between the signal and idler fields. A type-I, PPKTP second-harmonic generation (SHG) crystal temperature stabilized at 68° C is used to generate up to 500 mW of 470.98 nm blue light by pumping with a mode-locked Ti:Sapphire Tsunami laser (fundamental beam) operating at 76 MHz repetition rate with 50 ps pulse width. An optical schematic of the down-conversion source is given in Fig. 1(a).

The two outward facing sides of the PPKTP and KTP crystals are HR coated at 942 nm with reflectivities 99.8% and 90% respectively to form an asymmetric, single-sided, semi-confocal cavity with a finesse of 58 and a free-spectral range of ~8.9 GHz. The cavity is pumped from the high reflectance end of the PPKTP crystal and the down-converted photons are collected in the forward, collinear geometry through the KTP crystal using an f=60 mm doublet lens. The residual blue pump laser is isolated using dichroic filters and the signal and idler photons are differentiated using a polarizing beam cube. The cavity is locked using the Pound-Drever-Hall (PDH) technique [27] using the same fundamental laser with sidebands generated using a phase modulator (EOM). A chopper wheel operating at 50 Hz (See Fig. 1(a)) is used to alternate the locking of the laser and the collection of the down-converted photons. As the QD photons are collected in a single-mode fiber, the down-converted signal and idler photons are also collected using single-mode fibers. This ensures that the spatial mode of the down-conversion photons matches the QD photons for future applications (e.g., in a HOM interferometer).

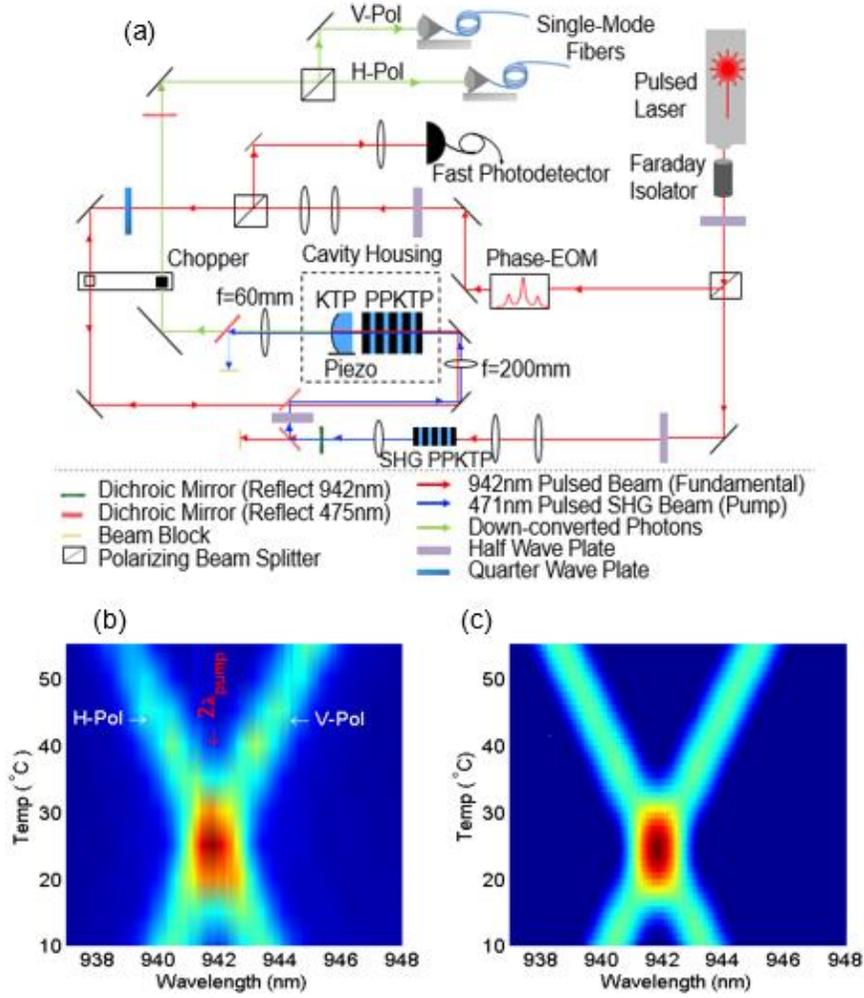

**Fig. 1** (a) Experimental schematic of the SHG and cavity-SPDC (b) Experimental data and (c) theoretical calculation for the spectrum of the down-converted photons as a function of PPKTP temperature. As theoretically predicted, the signal and idler spectrum are degenerate in wavelength with the fundamental beam at 27°C. The temperature bandwidth for degenerate operation is ~ 10°C.

Due to the birefringence nature of the crystals, the signal and idler fields acquire mismatched phases while propagating through them. With the *a priori* knowledge that the down-conversion bandwidth is much larger than the pump bandwidth (~6.3 GHz), the phase mismatch term is Taylor-expanded around the center frequencies assuming the zero pump bandwidth limit $(\Delta\omega_s = -\Delta\omega_i)$ and can be expressed as [28]

$$\Delta k\left(\omega_s + \Delta\omega_s, \omega_i + \Delta\omega_i, \omega_p\right) \approx \Delta k_0 + \frac{1}{c}\left[n_{gs} - n_{gi}\right]\Delta\omega_{si} \qquad (1)$$

where the first term, $\Delta k_0 = k_p(\omega_p) - k_s(\omega_s) - k_i(\omega_i) - 2\pi/\Lambda$, is the wave-vector mismatch between the pump, signal, and idler fields at single frequencies along with the contribution from the poling of the PPKTP crystal ($\Lambda = 33.25$ μm) designed to set $\Delta k_0 = 0$ at an appropriate crystal temperature. The temperature tuning is performed by mounting the PPKTP and KTP crystals on top of two separate thermo-electric coolers (TEC) (manufactured by TE Technology) that can adjust the crystals' temperatures from 10° C to 60° C. They are temperature stabilized within a few mK accuracy using two independent PID controllers (LFI-3751 manufactured by Wave Electronics). The PPKTP crystal temperature is tuned to 27° C to achieve degenerate operation at 941.85 nm. Similarly, the second term arises from the group index mismatch between the signal ($n_{gs}$) and idler ($n_{gi}$) fields, where $\Delta\omega_{si}$ is the bandwidth of the down-converted fields that determines the gain bandwidth of the cavity. Assuming the down-conversion intensity of the PPKTP crystal is proportional to $Sinc^2\left(\Delta k\left(\omega_s,\omega_i\right)L/2\right)$, the FWHM bandwidth (cavity gain) can be estimated by solving $\Delta k\left(\omega_s,\omega_i\right)L/2 = 1.39$ [28],

$$\Delta\omega_{si} = 2\frac{2\times 1.39\ c}{2\pi L \left|n_{gs}-n_{gi}\right|},\qquad(2)$$

where for a 5 mm long PPKTP crystal length (L) the down-conversion bandwidth is 540 GHz at 941.85 nm emission wavelength.

Fig. 1(b) is the spectrum of the signal and idler photons measured as a function of the PPKTP crystal temperature and Fig. 1(c) is the theoretical prediction calculated using the Sellemeir equations [29-31]. The measured FWHM of the signal photon is 610±10 GHz compared to the theoretical prediction of 540 GHz. The measured 'X' tuning behavior is a signature of the type-II phase matching and, as the data indicates, the fields are degenerate at 27° C, consistent with theoretical calculations [29-31].

### 3. First- and second-order coherence of cavity-SPDC photons

As the down-conversion bandwidth of the PPKTP crystal is significantly larger than the FSR of the cavity, a large number of cavity modes are occupied by the signal and idler fields, making the source highly multimode in spectrum [32]. For a continuous-wave pump laser with spectral bandwidth smaller than the FSR mismatch between the signal and idler modes of the cavity, down-conversion happens only at the spectral regions that have overlapping modes for the cavity, signal and idler fields. This results in the suppression of the unwanted emission modes [33-35]. However, for the emission time of the SPDC photons to have a small timing jitter and be well synchronized with other sources such as QD photons, the cavity is pumped with 50 ps fundamental pulses. This results in the pump bandwidth (6.35 GHz) to be significantly larger than the FSR mismatch between the signal and idler cavity modes (~0.2 GHz), resulting in the occupation of all the cavity modes within the down-conversion bandwidth. For a single spectral mode operation, the KTP crystal is tuned to obtain a double-resonance between the signal and idler fields degenerate with the QD photons and the output fields are filtered with a 6 GHz etalon centered at the QD emission. An intra-cavity etalon might be able to actively suppress these other modes while enhancing the count rate if made resonant with the central mode with an FSR large compared to the down conversion bandwidth. This is important as the spectral mode overlap between the cavity-SPDC photons and QD photons is essential to achieve a high HOM visibility.

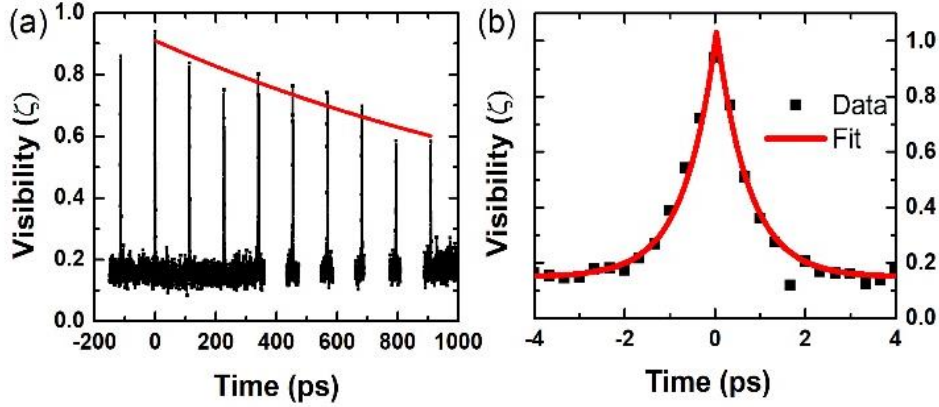

**Fig. 2**(a) Coherence time of the heralded cavity-SPDC photons measured using a Michelson interferometer. The interference visibility is calculated using the fringe contrast obtained while scanning the relative time difference between the two interferometer paths. The peaks are separated by 1/FSR = 113.6 ps of the cavity and the envelope decays gives the coherence time of the photons $\tau_c$= 1681±245 ps. The missing data are due to the limitation in the scan steps. (b) Expanded interference visibility data centered at zero-delay. The data is fitted with exponential decays with lifetime 1.4±0.2 ps corresponding to the SPDC down-conversion bandwidth.

The multiple cavity modes are evident in the data plotted in Fig. 2(a) with multiple interference peaks with decaying interference visibility obtained from the first-order coherence measurement ($g^{(1)}(\tau)$) [32]. The heralded signal photons are sent to a Michelson interferometer where the photons transmitted at an output port of the interferometer are counted and the interference visibility is calculated using the fringe contrast ($\zeta$) obtained while scanning the relative time difference between the two interferometer paths,

$$\zeta = \frac{\overline{I}_{max} - \overline{I}_{min}}{\overline{I}_{max} + \overline{I}_{min}} = \left|g^{(1)}(\tau)\right|. \quad (3)$$

The interference peaks are separated by 1/FSR=113.6 ps of the cavity. The envelope decay gives the coherence time of the field $\tau_c$=1681±245 ps, where the first order coherence time is twice the cavity lifetime of the cavity photons within the error bars. The slow modulation of the decay envelope is due to the higher-order transverse cavity modes; these appear as an additional small peak between the FSR of the cavity [36, 37] and can be seen in the high-resolution spectrum plotted data in Fig. 5. Fig. 2(b) is the expanded interference visibility data centered at zero delay. The data is fitted with exponential decays with lifetime 1.4±0.2 ps corresponding to the inverse of the SPDC down-conversion bandwidth.

To verify that the heralded photon source is non-classical and exhibits sub-Poissonian statistics, a Handbury Brown-Twiss (HBT) type three-detector intensity correlation measurement can be performed, where the correlation function can be expressed in terms of the signal and idler electric-field operators ($E_s$, $E_I$) as [19, 32, 38-40]

$$g^{(3)}_{s,s|I}(t,\tau_1,\tau_2) = \frac{\langle E_I^\dagger(t) E_s^\dagger(t+\tau_1) E_s^\dagger(t+\tau_2) E_s(t+\tau_2) E_s(t+\tau_1) E_I(t)\rangle}{\langle E_I^\dagger(t) E_I(t)\rangle \langle E_s^\dagger(t) E_s(t)\rangle^2}. \quad (4)$$

If a source emits a single pair of signal and idler photons at a given shot of the experiment, the probability of all three detectors firing simultaneously is zero, thus $g^{(3)}_{s,s|I}(\tau=0)$ is an important figure of merit that can be used to quantify the multi-photon emission probability of the source. A heralded intensity correlation measurement is performed by detecting an idler photon by detector $D_I$ and sending the signal photon to a 50:50 non-polarizing beam-splitter and detecting the transmitted and reflected photons with detectors $D_2$ and $D_3$, where all the photons are collected into single-mode fibers. The single and joint detection count is recorded using a coincidence counter from which the heralded intensity correlation of the signal photon is calculated as [32, 38-41]

$$g_{ss|I}^{(3)}(0) = \frac{N_1 N_{123}}{N_{12} N_{13}}. \tag{5}$$

Where $N_{123}$ is the number of triple coincidence counts between all three detectors, $N_{12}$ is the number of joint detection events between detectors 1 and 2, $N_{13}$ is the number of joint detection events between detectors 1 and 3 and $N_1$ is the total number of heralding photons measured by detector 1. All the single and coincidence counts are recorded within 3 ns of the heralding idler event, well beyond the coherence time of the down-conversion photons.

The result from the $g_{ss|I}^{(3)}$ measurements performed as a function of the pump power are plotted in Fig. 3. As indicated by the data, the signal photon exhibits sub-Poissonian statistics with $g_{ss|I}^{(3)} = 0.071 \pm 0.035$ for 5 mW of pump power, making it an excellent source of heralded single-photons. For a reasonably high pump power (<65 mW), the source still exhibits a sub-Poissonian behavior with $g_{ss|I}^{(3)} < 0.5$. As the pump power increases, the contribution of the higher order terms – which represent the multi-photon emission rates – become more significant in the down-conversion wavefunction. This results in a coincidence-count/single-photon-purity tradeoff and limits the use of the source for applications that put stringent requirements for these two criteria.

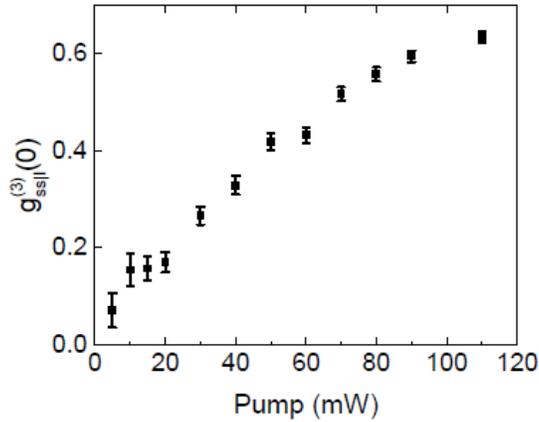

**Fig. 3.** Photon statistics of a heralded cavity-SPDC photon as a function of the excitation pump power obtained with three-detector correlation measurements, where the first detector heralds the presence of a signal photon and the signal photon is sent to a HBT setup to verify its single photon nature. The plotted data is obtained by measuring the triple coincidence counts within a 3 ns window of the heralding event as a function of pump power. The data indicates that for low pump power, $g_{ss|I}^{(3)}$ is close to zero and as the pump-power is increased the probability of higher-pair generation increases, resulting in the reduction in the $g_{ss|I}^{(3)}$ value. Nonetheless, the $g_{ss|I}^{(3)}$ is still below the classical limit (0.5) for a reasonably high pump power up to 65 mW.

## 4. Interfacing cavity-SPDC photons with a single quantum dot

Single semiconductor QD nano-structures formed on InAs/GaAs material are well known to exhibit atom-like behavior and form effective two-level systems under resonant optical excitation [42]. An optically generated exciton state along with a single electron trapped inside a QD forms a trion state and is shown to generate a spin-photon entanglement source [43-45]. The ground state of the trion (consisting of a single electron) forms a matter qubit [46-48], an essential building block for a quantum network. Here, to demonstrate that the mode engineered cavity-SPDC source can be coupled with such a matter qubit, we resonantly excite a single trion state centered at 941.85 nm using the cavity-SPDC source.

The QDs are embedded in an asymmetric distributed Bragg reflector (DBR) cavity with a low Q factor of ~90. The DBR structure is designed to enhance the emission in the direction away from and generally in the direction perpendicular to the substrate resulting in a significant increase in detected emission [49, 50]. The sample is grown with a PIN-diode structure that allows selective charging of a single QD to form a trion state. In addition, the applied bias voltage allows the transition energy of the QD to be Stark-shifted by several wavenumbers, allowing us to perform voltage dependent resonance fluorescence measurements [51, 52]. The QD sample is cooled to 5.6 K using a liquid helium bath cryostat.

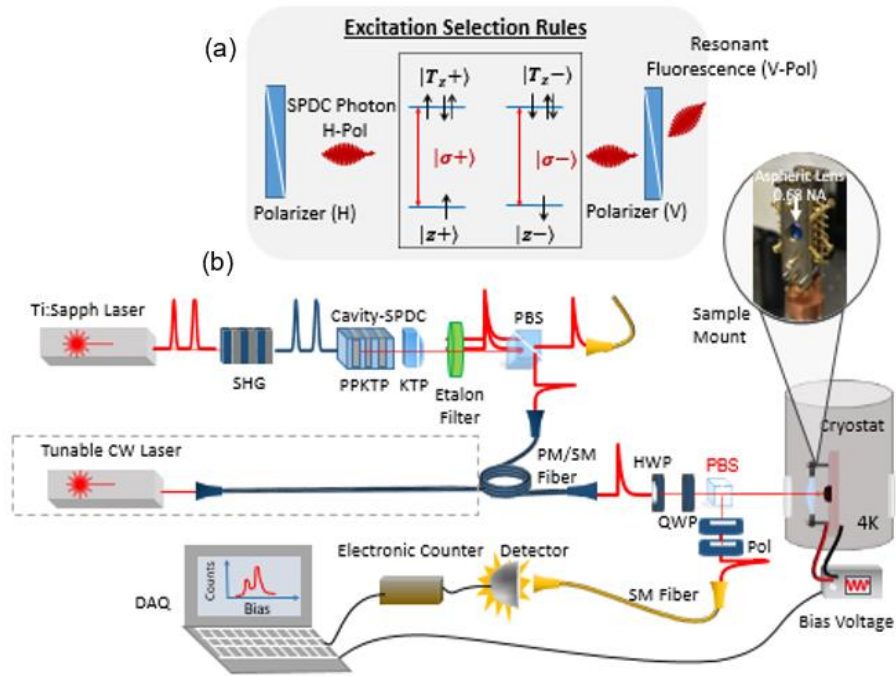

**Fig. 4(a)** Optical selection rule of a single QD (trion state) at zero external magnetic field. As the selection rule indicates, when a single trion state is in resonance with the excitation field, it emits photons with σ+ and σ− polarizations. By counting emitted photons orthogonal to the excitation light, we can verify the resonant interaction between the incident field and the single QD. (b) Experimental setup for direct excitation of a single QD with the SPDC photons. The experiment is performed in a dark-field microscopy setup where the incident SPDC photons are focused to a single QD with a high NA (0.65) aspheric lens and the same lens is used to collect the emitted photons. The dashed box is a removable setup used to find a single QD in the study. Once a single QD is identified at the wavelength degenerate with the down-conversion photons, the fiber connecting the CW-laser to the setup is disconnected and connected to the photons collected from the cavity-SPDC source.

The schematic of the optical experiment for exciting and detecting the photons emitted by a single QD is given in Fig. 4, with Fig. 4(a) showing the optical transition selection rules of a single trion state in the absence of an external magnetic field. As seen in the figure, a single trion state forms two degenerate two-level systems that are optically coupled by σ+ and

σ− polarized light [52, 53]. Thus, by exciting the QD with horizontally polarized light and collecting vertically polarized photons, we can verify the resonant excitation of the trion state.

The optical experiment is performed in a dark-field microscopy setup [54, 55] where the incident light is focused to a single QD with a high NA (0.65) aspheric lens and the same lens is used to collect the emitted photons. Using a narrowband continuous-wave laser, a single trion state is identified by performing resonance Rayleigh scattering measurement at a fixed non-zero dc bias point [55]. Then the transition energy of the QD is detuned relative to the incident laser by scanning the dc bias voltage while keeping the CW-laser at the fixed frequency. The bias voltage tunes the resonance frequency of the QD due to the Stark effect. Fig. 5(a) shows the resonant fluorescence spectrum of a single trion state measured by exciting the QD with a narrow bandwidth CW laser. The QD's linewidth is measured to be 730±15 MHz, broadened beyond its radiative lifetime limit, which is attributed to spectral diffusion due to charge noise in the sample. In Figs. 4(a), 4(b), and 4(d) the center frequency is detuned from 941.84307 nm.

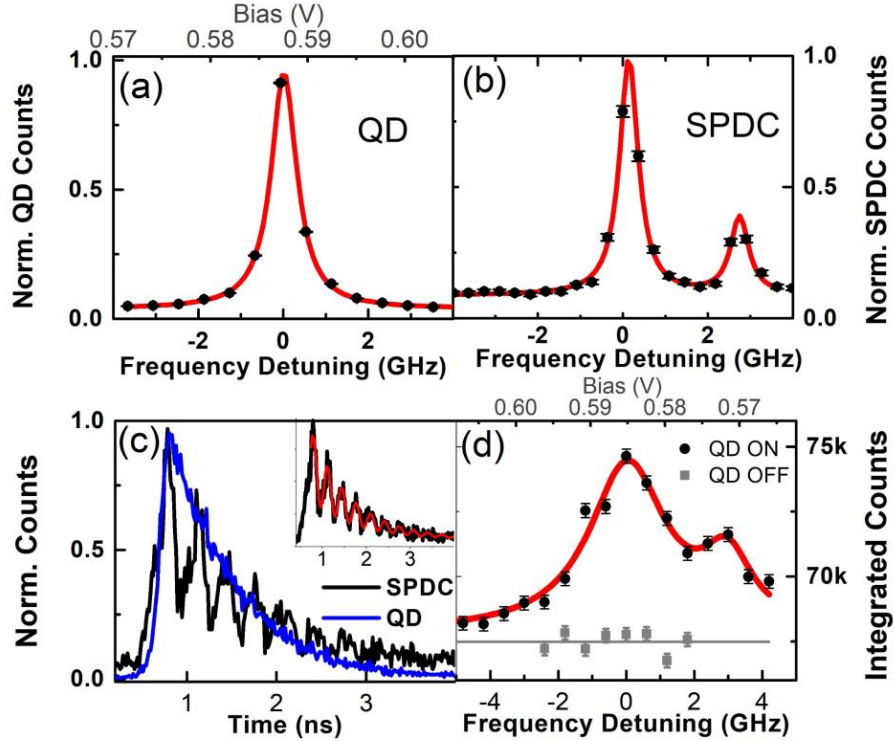

**Fig. 5**. (a) Resonant Rayleigh scattering spectrum of a single trion state measured by exciting the QD with a separate narrow bandwidth CW-laser. Rather than scanning the frequency of the incoming radiation, we sweep the excitation spectrum of the QD through the fix excitation frequency by tuning the absorption resonance via the Stark effect induced by the applied bias. In plot a), b), and d) the center frequency is detuned from 941.84307 nm. (b) SPDC spectrum measured using an external scanning Fabry-Pérot etalon, the center frequency component corresponds to the $TEM_{00}$ mode of the cavity and the peak to the right separated by 2.8±0.2 GHz corresponds to the higher order transverse modes of the cavity. (c) time-tagged photon emissions from a single trion state (blue) and SPDC signal photons (black) with lifetimes 751±11 ps and 932±50 ps respectively, which correspond to the natural linewidth $\gamma_2/(2\pi)$ of 212 MHz and 171 MHz. The inset is the SPDC photon lifetime fitted with a 3.06±0.1 GHz oscillation arising due to the beating between the cavity modes, consistent with the spectral profile obtained in (b). (d) SPDC photons scattered by a single QD as the QD is brought in and out of resonance with the SPDC photons. The two peaks separated by 3 GHz are mapped to the photons scattered by the QD. The gray dots are the data with the QD bias voltage off, which turns the QD transition off. This results in the drop of the scattered photon counts to the background level.

Fig. 5(b) is the spectrum of the cavity-SPDC signal photons (after the filter etalon) used to excite the QD. The spectrum of the signal photons after the filter etalon is measured by using a pressure-tuned etalon (FSR = 45 GHz, Bandwidth = 400 MHz) where the x-axis corresponds to the Fabry–Pérot etalon detuning from the center wavelength of the QD (941.84307 nm) and the measured linewidth is limited by the bandwidth of the etalon. The center peak is due to the $TEM_{m=0;n=0}$ mode of the cavity and the smaller peak on the right is due to a higher order cavity mode [36, 37], which are separated by 2.8±0.2 GHz. Due to the finite bandwidth of the filter etalon, both peaks are used to excite the QD. Fig. 5(c) is the temporal profile of the cavity-SPDC signal photons (black) and the QD photons (blue) both obtained using the same Tsunami laser and building time-tagged emission by syncing the emission time with the excitation laser pulse centered at 941.85 nm. The QD in the study has a lifetime of 751±11 ps and the SPDC has a lifetime of 932±50 ps with a relative mismatch of 21.5%. The cavity-SPDC lifetime can be tuned between 800 ps to 1200 ps by changing the cavity length. The inset shows the oscillation of the temporal profile of the cavity-SPDC photons due to the beating between the two cavity modes. The red curve is an exponentially decaying curve with sinusoidal oscillation fit to the integrated counts of cavity-SPDC signal photons right before they sent to interact with the QD. The oscillation frequency is 3.08±0.1 GHz, consistent with the spectral profile of signal photons given in Fig. 5(b).

To show resonant coupling between the cavity-SPDC photons and a single trion state, the CW-laser (dashed box in Fig. 4) exciting the QD is replaced with the signal photons collected from the cavity-SPDC source. For this measurement, the cavity-SPDC source is excited with 40 mW of blue light and the collected signal photons are sent through the QD setup. The polarization of the signal photons is rotated with a quarter and a half-wave plate before the PBS to correct for any polarization rotation while traveling through the fiber and thus maximize the transmission through the beam-splitter. The QD is excited with ~60,000 signal photons per second. As described previously, when linearly polarized input cavity-SPDC photons are in resonance with the QD transition energy, the QD is excited and emits circularly polarized single photons. By collecting the photons that are orthogonally polarized to the incident photons, we can ensure the collected photons are emitted by the QD, thus verify a direct excitation of the QD by SPDC photons.

We detected the emitted photons with high-efficiency superconducting nanowire detectors while changing the resonance energy of the QD transition with the Stark-shift effect. The black dots in Fig. 5(d) correspond to the total photon count integrated for 10 minutes at each bias voltage. And the gray dots correspond to the background counts when the QD is turned off by removing the bias. When the QD is turned off, the counts are primarily the detector dark counts 110±10, corresponding to 66,000 dark counts over a 10 min. integration time (grey dots Fig. 5(d)). As we turn the QD on and tune the bias voltage, the two different frequency components of the cavity-SPDC photons separated by 3 GHz come in and out of resonance with the QD transition. As a result, more photons are scattered and detected at the resonances. As one would expect from the Stark shift of the trion, the higher frequency peak shows up at lower bias, consistent with the measured data. All the key features of the SPDC photons are mapped on to the direct excitation data seen in Fig. 5(d). The red curve is a fit to the data with two peaks that are split by 3 GHz, consistent with the cavity-SPDC spectral modes used to excite the QD. The broadening of the linewidth is due to the cavity-SPDC drift over time, which is not locked for this measurement in order to increase the count rate that otherwise would be blocked by the optical chopper. With this data, we have successfully demonstrated a cascaded system where a single QD is excited with spectral engineered photons emitted from a cavity-SPDC source.

## 5. Conclusions

In summary, we have built a polarization-correlated photon pair source using cavity-SPDC that matches the bandwidth, wavelength, and temporal profile of an InAs/GaAs QDs centered at 941.85 nm. We showed that the cavity-SPDC source exhibits a sub-Possionian behavior for a reasonably high pump power of up to 65 mW. By directly exciting a single QD with cavity-SPDC photons, we have demonstrated the direct coupling between the SPDC photon and a QD. The current work could be immediately extended to show a heralded transfer of a photonic qubit generated from the frequency modulated cavity-SPDC photons using an external modulator [55] to the spin ground states of a charge QD through a direct absorption measurement [12, 14]. In addition, such photonic qubits could be teleported to the spin state of a single QD by interfering it with the photons of a spin-photon entangled state [56]. This would require a high HOM interference visibility between the cavity-SPDC and QD photons which demands a strict spectral and temporal mode overlap between the two photons [57]. The observed mismatch in the SPDC lifetime caused by the beat frequency can be removed by using an intra-cavity etalon that has the same bandwidth as a single mode of the cavity and an FSR larger than the down-conversion bandwidth. Such experiments would be of interest for building a quantum memory or a quantum repeater for realizing a cascaded quantum link between disparate systems.

**Funding.** This work is supported in part by NSF (PHY 1413821), AFOSR (FA9550-09-1-0457), ARO (W911NF-08- 1-0487, W911NF-09-1-0406/Z855204), DARPA (FA8750- 12-2-0333), and the Office of Naval Research.

**Acknowledgment.**
We would like to thank Jian Wang and Alex Burgers for working on the earlier stage of the SPDC source design, Aaron Ross for building the scanning Fabry-Pérot étalon and Marta Luengo-Kovac for carefully reading the manuscript. MG would like to thank the Univ. of Illinois for support and hospitality during his sabbatical visit.

## References

1. H. J. Kimble, "The Quantum Internet," Nature, **453**(1023), 1023–1030 (2008).
2. C. Jones, D. Kim, M. T. Rakher, P. G. Kwiat, and T. D. Ladd, "Design and analysis of communication protocols for quantum repeater networks," New J. Phys., **18**, 083015 (2016).
3. D. Leibfried, E. Knill, S. Seidelin, J. Britton, R. B. Blakestad, J. Chiaverini, D. B. Hume, W. M. Itano, J. D. Jost, C. Langer, R. Ozeri, R. Reichle, and D. J. Wineland, "Creation of a six-atom `schrodinger cat' state," Nature, **438**(7068), 639–642 (2005).
4. D. Kim, S. G. Carter, A. Greilich, A. S. Bracker, and D. Gammon, "Ultrafast optical control of entanglement between two quantum-dot spins," Nat. Phys., **7**, 223–229 (2010).
5. D. L. Moehring, P. Maunz, S. Olmschenk, K. C. Younge, D. N. Matsukevich, L.M. Duan, and C. Monroe, "Entanglement of single-atom quantum bits at a distance," Nature, **449**(7158), 68–71 (2007).
6. W A. Delteil, Z. Sun, W.-B. Gao, E. Togan, S. Faelt, and A. Imamoğlu, "Generation of heralded entanglement between distant hole spins," Nat. Phys. **12**, 218–223 (2016).
7. E. Flagg, A. Muller, S. Polyakov, A. Ling, A. Migdall, and G. Solomon, "Interference of single photons from two separate semiconductor quantum dots," Phys. Rev. Lett., **104**(13), 137401 (2010).
8. S. Ritter, C. Nlleke, C. Hahn, A. Reiserer, A. Neuzner, M. Uphoff, M. Mcke, E. Figueroa, J. Bochmann, and G. Rempe, "An elementary quantum network of single atoms in optical cavities," Nature, **484**(7393), 195-200 (2012).
9. Y. L. A. Rezus, S. G. Walt, R. Lettow, A. Renn, G. Zumofen, S. Gtzinger, and V. Sandoghdar, "Single-photon spectroscopy of a single molecule," Phys. Rev. Lett., **108**, 093601 (2012).
10. N. Kalb, A. Reiserer, S. Ritter, and G. Rempe, "Heralded storage of a photonic quantum bit in a single atom," Phys. Rev. Lett., **114**, 220501 (2015).
11. N. Piro, F. Rohde, C. Schuck, M. Almendros, J. Huwer, J. Ghosh, A. Haase, M. Hennrich, F. Dubin, and J. Eschner, "Heralded single-photon absorption by a single atom," Nat. Phys., **7**, 17-20 (2011).
12. A. Delteil, Z. Sun, S. Fält, and A. Imamoglu, " Realization of a cascaded quantum system: Heralded absorption of a single photon qubit by a single-electron charged quantum dot," Phys. Rev. Lett., **118**, 177401 (2017).
13. J. I. Cirac, P. Zoller, H. J. Kimble, and H. Mabuchi, " Quantum state transfer and entanglement distribution among distant nodes in a quantum network," Phys. Rev. Lett., **78**, 3221 (1997).
14. D. Pinotsi and A. Imamoglu, "Single photon absorption by a single quantum emitter," Phys. Rev. Lett., **100**, 093603 (2008).
15. H. M. Meyer, R. Stockill, M. Steiner, C. L. Gall, C. Matthiesen, E. Clarke, A. Ludwig, J. Reichel, M. Atature, and M. Khl, " Direct photonic coupling of a semiconductor quantum dot and a trapped ion," Phys. Rev. Lett., **114**, 123001 (2015).

16. G. Schunk, U. Vogl, D. V. Strekalov, M. Förtsch, F. Sedlmeir, H.G.L. Schwefel, M. Göbelt, S. Christiansen, G. Leuchs, and C. Marquardt, "Interfacing transitions of different alkali atoms and telecom bands using one narrowband photon pair source," Optica **2**(9), 773-778 (2015).
17. V. Leong, M. A. Seidler, M. Steiner, A. Cerè, and C. Kurtsiefer, "Time-resolved scattering of a single photon by a single atom," Nat. Comm. **7**, 13716 (2016).
18. S. V. Polyakov, A. Muller, E. B. Flagg, A. Ling, N. Borjemscaia, E. V. Keuren, A. Migdall, and G. S. Solomon, "Coalescence of single photons emitted by disparate single-photon sources: The example of InAs quantum dots and parametric down-conversion sources," Phys. Rev. Lett., **107**, 157402 (2011).
19. F. Wolfgramm, Y. A. d. I. Astiz, F. A. Beduini, A. Cere, and M. W. Mitchell, "Atom-resonant heralded single photons by interaction-free measurement," Phys. Rev. Lett., **106**, 053602 (2011).
20. P. G. Kwiat, K. Mattle, H. Weinfurter, A. Zeilinger, A. V. Sergienko, and Y. Shih, "New high-intensity source of polarization-entangled photon pair," Phy. Rev. Lett., **75**, 4337 (1995).
21. F. Kaneda, K. Garay-Palmett, A. B. U'Ren, and P. G. Kwiat, "Heralded single-photon source utilizing highly nondegenerate, spectrally factorable spontaneous parametric downconversion," Opt. Express, **24**(10), 10733-10747 (2016).
22. Z. Y. Ou and Y. J. Lu, "Cavity enhanced spontaneous parametric downconversion for the prolongation of correlation time between conjugate photons.," Phys. Rev. Lett., 83, 2556 (1999).
23. M. Scholz, L. Koch, and O. Benson, "Statistics of narrow-band single photons for quantum memories generated by ultrabright cavity-enhanced parametric down-conversion," Phys. Rev. Lett., **102**, 063603 (2009).
24. S. A. Aljunid, G. Maslennikov, Y. Wang, H.L. Dao, V. Scarani, and C. Kurtsiefer, " Excitation of a single atom with exponentially rising light pulses," Phys. Rev. Lett, **111**, 103001 (2013).
25. B. Srivathsan, G.K. Gulati, A. Cerè, B. Chng, and C. Kurtsiefer, "Reversing the temporal envelope of a heralded single photon using a cavity," Phys. Rev. Lett., **113**, 163601 (2014).
26. M. Stobińska, G. Alber, and G. Leuchs, "Perfect excitation of a matter qubit by a single photon in free space," EPL, **86**(1), 14007 (2009).
27. R. W. P. Drever, J. L. Hall, F. V. Kowalski, J. Hough, G. M. Ford, A. J. Munley, and H. Ward, "Laser phase and frequency stabilization using an optical resonator," Appl. Phys. B, **31**(2), 97-105 (1983).
28. Z.-Y. J. Ou, *Multi-Photon Quantum Interference* (Springer, 2007).
29. Z. Zeng, H. Shen, H. Xu, Y. Zhou, C. Huang, and D. Shen, "Measurement of refractive indices and thermal refractive index coefficients of the Ti:Mg:LiNbO3 crystal," J. Synth. Cryst., **16**(3), 551-553 (1987).
30. H. Y. Shen, Y. P. Zhou, W. X. Lin, Z. D. Zeng, R. R. Zeng, G. F. Yu, C. H. Huang, A. D. Jiang, S. Q. Jia, and D. Z. Shen, " Second harmonic generation and sum frequency mixing of dual wavelength Nd:YALO3 laser in flux grown KTiOPO4 crystal," IEEE J. Quantum Electron., **28**, 41-55 (1992).
31. K. Kato and E. Takaoka, "Sellmeier and thermo-optic dispersion formulas for KTP," Appl. Optics, **41**(24), 5040-5044 (2002).
32. M. Scholz, L. Koch, and O. Benson, "Analytical treatment of spectral properties and signal–idler intensity correlations for a double-resonant optical parametric oscillator far below threshold," Opt. Commun., **282**, 3518-3523 (2009).
33. C.-S. Chuu, G. Y. Yin, and S. E. Harris, "A miniature ultrabright source of temporally long, narrowband biphotons," Appl. Phys. Lett., **101**, 051108 (2012).
34. P.-J. Tsai and Y.-C. Chen, "Ultrabright, narrow-band photon-pair source for atomic quantum memories," QST, **3**(3), 034005 (2018).
35. K.-H. Luo, H. Herrmann, S. Krapick, B. Brecht, R. Ricken, V. Quiring, H. Suche, W. Sohler, and C. Silberhorn, "Direct generation of genuine single-longitudinal-mode narrowband photon pairs," New J. Phys., **17**(7), 073039 (2015).
36. G. D. Boyd and H. Kogelnik, "Generalized confocal resonator theory," Bell Syst. Tech. J., **41** (4), 1347 - 1369 (1962).
37. J. P. Goldsborough, "Beat frequencies between modes of a concave-mirror optical resonator," Appl. Optics, **3**(2), 267-275 (1964).
38. P. Grangier, G. Roger, and A. Aspect, "Experimental evidence for a photon anticorrelation effect on a beam splitter: A new light on single-photon interferences," EPL, **1**(4), 173 (1986).
39. E. Bocquillon, C. Couteau, M. Razavi, R. Laflamme, and G. Weihs, "Coherence measures for heralded single-photon sources," Phys. Rev. A, **79**, 035801 (2009).

40. S. Fasel, O. Alibart, S. Tanzilli, P. Baldi, A. Beveratos, N. Gisin, and H. Zbinden, "High-quality asynchronous heralded single-photon source at telecom wavelength," New J. Phys., **6**(1), 163 (2004).
41. M. Rambacha, A. Nikolova, T. J. Weinhold, and A. G. White, "Sub-megahertz linewidth single photon source," APL Photonics, **1**(9), 096101 (2016).
42. T. H. Stievater, X. Li, D. G. Steel, D. Gammon, D. S. Katzer, D. Park, C. Piermarocchi, and L. J. Sham, "Rabi oscillations of excitons in single quantum dots," Phys. Rev. Lett., **87**, 133603 (2001).
43. W. B. Gao, P. Fallahi, E. Togan, J. Miguel-Sanchez, and A. Imamoglu, " Observation of entanglement between a quantum dot spin and a single photon," Nature, **491**, 426–430 (2012).
44. K. D. Greve, L. Yu, P. L. McMahon, J. S. Pelc, C. M. Natarajan, N. Y. Kim, E. Abe, S. Maier, C. Schneider, M. Kamp, S. Höfling, R. H. Hadfield, A. Forchel, M. M. Fejer, and Y. Yamamoto, "Quantum-dot spin–photon entanglement via frequency downconversion to telecom wavelength," Nature, **491**, 421–425 (2012).
45. J. R. Schaibley, A. P. Burgers, G. A. McCracken, L.-M. Duan, P. R. Berman, D. G. Steel, A. S. Bracker, D. Gammon, and L. J. Sham, "Demonstration of quantum entanglement between a single electron spin confined to an InAs quantum dot and a photon," Phys. Rev. Lett., **110**, 167401 (2013).
46. D. Press, T. D. Ladd, B. Zhang, and Y. Yamamoto, "Complete quantum control of a single quantum dot spin using ultrafast optical pulses," Nature, **456**, 218-221 (2008).
47. M. Atatüre, J. Dreiser, A. Badolato, A. Högele, K. Karrai, and A. Imamoglu, "Quantum-dot spin-state preparation with near-unity fidelity," Science, **312**(5773), 551-553 (2006).
48. X. Xu, Y. Wu, B. Sun, Q. Huang, J. Cheng, D. G. Steel, A. S. Bracker, D. Gammon, C. Emary, and L. J. Sham, "Fast spin state initialization in a singly charged InAs-GaAs quantum dot by optical cooling," Phys. Rev. Lett., **99**, 097401 (2007).
49. H. Benisty, H. D. Neve, and C. Weisbuch, "Impact of planar microcavity effects on light extraction-part I: basic concepts and analytical trends," IEEE J. Quantum Electron., **34**(9), 1612 - 1631 (1998).
50. O. Gazzano and G. S. Solomon, "Toward optical quantum information processing with quantum dots coupled to microstructures," J. Opt. Soc. Am. B, **33**(7), C160-C175 (2016).
51. B. Alén, F. Bickel, and K. Karrai, "Stark-shift modulation absorption spectroscopy of single quantum dots," Appl. Phys. Lett., **83**, 2235 (2003).
52. X. Xu, B. Sun, E. D. Kim, K. Smirl, P. R. Berman, D. G. Steel, A. S. Bracker, D. Gammon, and L. J. Sham, "Single charged quantum dot in a strong optical field: absorption, gain, and the ac-Stark effect," Phys. Rev. Lett., **101**, 227401 (2008).
53. M. E. Ware, E. A. Stinaff, D. Gammon, M. F. Doty, A. S. Bracker, D. Gershoni, V. L. Korenev, S. C. Badescu, Y. Lyanda-Geller, and T. L. Reinecke, "Polarized Fine Structure in the Photoluminescence Excitation Spectrum of a Negatively Charged Quantum Dot," Phys. Rev. Lett., **95**, 177403 (2005).
54. A. V. Kuhlmann, J. Houel, D. Brunner, A. Ludwig, D. Reuter, A. D. Wieck, and R. J. Warburton, "A dark-field microscope for background-free detection of resonance fluorescence from single semiconductor quantum dots operating in a set-and-forget mode," Rev. of Sci. Instrum., **84**, 073905 (2013).
55. U. Paudel, A. P. Burgers, D. G. Steel, M. K. Yakes, A. S. Bracker, and D. Gammon, "Generation of frequency sidebands on single photons with indistinguishability from quantum dots," Phys. Rev. A, **98**, 011802 (R) (2018).
56. W. Gao, P. Fallahi, E. Togan, A. Delteil, Y. Chin, J. Miguel-Sanchez, and A. Imamoğlu, "Quantum teleportation from a propagating photon to a solid-state spin qubit," Nat. Comm., **4**, 2744 (2013).
57. B. Kambs and C. Becher, "Limitations on the indistinguishability of photons from remote solid state sources," New. J. Phys., **20**, 115003 (2018).